\documentclass[aps,prd,nofootinbib,floats,floatfix,preprintnumbers,groupedaddress,twocolumn]{revtex4}
\usepackage{bm}
\usepackage{latexsym}
\usepackage{dcolumn}
\usepackage{amsmath,amsfonts,amssymb}
\usepackage{graphicx,epsfig}
\usepackage{color}
\def\beq{\begin{equation}}
\def\eeq{\end{equation}}
\def\bea{\begin{eqnarray}}
\def\eea{\end{eqnarray}}
\def\benu{\begin{enumerate}}
\def\eenu{\end{enumerate}}
\def\nn{\nonumber}

\def\l{\left}
\def\r{\right}

\reversemarginpar

\def\DM{\mathrm{d}}
\def \ub {u}
\begin{document}

\title{Box of Ideal Gas in Free Fall}% in Curved Spacetime}

\author{
Dawood Kothawala \footnote{Current Affiliation: Dept. of Physics, I.I.T. Madras, Chennai 600036, Tamil Nadu, India.}
}
\email{dawood@physics.iitm.ac.in}

\affiliation{Department of Mathematics and Statistics, University of New Brunswick, Fredericton, NB, Canada E3B 5A3.
}

\date{\today}

\begin{abstract}
We study the \textit{quantum} partition function of non-relativistic, ideal gas in a (non-cubical) box falling freely in arbitrary curved spacetime with centre $4$-velocity $u^a$. When perturbed energy eigenvalues are properly taken into account, we find that corrections to various thermodynamic quantities include a very specific, sub-dominant term which is independent of \textit{kinematic} details such as box dimensions and mass of particles. This term is characterized by the dimensionless quantity, $\Xi=R_{{\hat 0}{\hat 0}} \Lambda^2$, where $R_{{\hat 0}{\hat 0}}=R_{ab} u^a u^b$ and $\Lambda=\beta \hbar c$, and, quite intriguingly, produces Euler relation of homogeneity two between entropy and energy -- a relation familiar from black hole thermodynamics. 
\end{abstract}

\maketitle
%%%%%%%%%%%%%%%%%%%%%%%%%%%%%%%%%%%%%%%%%%%%%%%%%%%%%%%%%%%%%%%%%%%%%%%%%%%%%%%%%%%%%%%%%%%%%
%\section*{Box of Ideal gas in free fall} \label{app:stat-mech-ideal-gas} 
%%%%%%%%%%%%%%%%%%%%%%%%%%%%%%%%%%%%%%%%%%%%%%%%%%%%%%%%%%%%%%%%%%%%%%%%%%%%%%%%%%%%%%%%%%%%%
\section{Introduction}%\noindent \textbf{\textit{Introduction}}: 
There have been several intriguing connections between gravity and thermodynamics discovered over the past few years (see \cite{paddy-newinsights} for a recent review), a better understanding of which necessitates study of thermal systems in presence of gravity. For example, a study of phase space available to thermal systems in the vicinity of spacetime horizons yields results which might be helpful to understand certain aspects of horizon entropy \cite{sanved}. It is indeed possible that certain features of black hole thermodynamics are simply features of standard thermodynamic systems when curvature of spacetime is accounted for in the analysis of the latter \cite{bei-lok-hu}. Now, we know that black hole thermodynamics is inherently quantum mechanical in origin, and hence one may not learn anything ``drastically" new simply from curvature corrections to standard systems using \textit{classical} statistical mechanics; at best, classical analysis can yield terms representing tidal forces (needed to hold the box together) etc. It is more useful to ask whether a \textit{quantum mechanical} calculation can give any new information, which is the question we hope to address in this note in the context of one of the simplest thermodynamic systems -- a box of ideal gas. We consider such a box of ideal gas in an arbitrary curved spacetime, with its center freely falling along a geodesic with 4-velocity $\bm \ub$, and compute corrections to the partition function due to spacetime curvature. And indeed, we find that all thermodynamic quantities acquire a specific correction term (besides others) which is independent of system details such as box dimensions and mass of particles. This term is characterized by the dimensionless quantity $\Xi=R_{{\hat 0}{\hat 0}} \Lambda^2$, where $R_{{\hat 0}{\hat 0}}=R_{ab} \ub^a \ub^b$ and $\Lambda=\beta \hbar c$. We highlight several features of this $\Xi$-contribution, in particular the fact that $S_{\Xi} = (1/2) \beta U_{\Xi}$, a relation which is familiar from black hole thermodynamics. It would be worthwhile to investigate other possible implications of these corrections from the viewpoint of thermodynamic aspects of gravity. 

We describe the set-up and relevant calculations in the next section; the perturbed free-particle energy eigenvalues are given in Eqs.~(6), and the final results are given in Eqs.~(9). In the final section, we conclude with a few remarks on implications of the result. In order-of-magnitude arguments, we will use $\mathcal{R}$ to denote typical magnitude of curvature tensor components. 

%%%%%%%%%%%%%%%%%%%%%%%%%%%%%%%%%%%%%%%%%%%%%%%%%%%%%%%%%%%%%%%%%%%%%%%%%%%%%%%%%%%%%%%%%%%%%
\section{Thermodynamics of box of ideal gas}%\vspace{0.2cm}
%\noindent \textbf{\textit{Thermodynamics of box of ideal gas}}: 
Consider a box of gas, whose center is following a timelike trajectory with $4$-velocity $\bm \ub$. The spacetime in a local neighbourhood of this trajectory can be constructed using Fermi normal coordinates (FNC), in which the metric takes the form \cite{misner-fnc}
\begin{eqnarray}
g_{{\hat 0} {\hat 0}} &=& - \l[ \l( 1 + \frac{a_{\mu} y^{\mu}}{c^2} \r)^2 + R_{{\hat 0} \mu {\hat 0} \nu} y^{\mu} y^{\nu} \r] + O(y^3)
\nn \\
\nn \\
g_{{\hat 0} \mu} &=& - \l(2/3\r) R_{{\hat 0} \rho \mu \sigma} y^{\rho} y^{\sigma} + O(y^3)
\nn \\
\nn \\
g_{\mu \nu} &=& \delta_{\mu \nu} - \l(1/3\r) \; R_{\mu \rho \nu \sigma} y^{\rho} y^{\sigma} + O(y^3)
\\ \nn
\end{eqnarray}
where Greek indices run over spatial coordinates, and $\bm a$ is the acceleration corresponding to $\bm \ub$. We shall consider a box whose center is falling freely, and therefore set $\bm a=0$. In these coordinates, $\bm \ub=\bm \partial_{{\hat 0}}$ (i.e., the original trajectory is simply $y^{\mu}=0$), and we define a ``box" as a confined region with flat ``coordinate" faces, i.e., $ -(L_x/2) \le x \le +(L_x/2)  $ and similarly for $y$ and $z$. (There would, of course, be tidal forces on box walls, necessitating some mechanism to keep them in place; this can always be taken care of and would not change the Hamiltonian of the particles inside the box.) This box is filled with an ideal gas at temperature $\beta^{-1}$. The Hamiltonian for the constituent particles can be written in FNC as $H = - p_{\hat 0}$, which can be motivated as follows. The action for a single particle can be written in covariant form as $\int p_i \DM x^i$. In FNC, the time coordinate $t_{\mathcal{P}}$ assigned to an event $\mathcal{P}$ in (a normal convex) neighbourhood of the trajectory is equal to the proper time $\tau$ at the point on the trajectory connected to $\mathcal{P}$ by a spacelike geodesic; i.e., $t_{\mathcal{P}}=\tau$. The action can therefore be split as $\int \l( p_{\hat 0} + p_{\mu} \DM x^{\mu}/\DM \tau  \r) \DM \tau$, which immediately identifies $H=- p_{\hat 0}$. Note that this Hamiltonian is {\it different} from $ H = - \bm p \bm \cdot \bm \ub^{\parallel} $, where $\bm \ub^{\parallel}$ is $\bm \ub$ parallel transported to the location of the particle. We can write out the Hamiltonian explicitly in terms of the metric coefficients using $\bm p \bm \cdot \bm p = -m^2$; this gives
\begin{eqnarray}
H = \frac{g^{{\hat 0} \mu} p_{\mu} c}{g^{{\hat 0 \hat 0}}} +  \sqrt{ \frac{g^{\mu \nu} p_{\mu} p_{\nu} c^2 + m^2 c^4}{-g^{{\hat 0 \hat 0}}} + \l( \frac{g^{{\hat 0} \mu} p_{\mu} c}{g^{{\hat 0 \hat 0}}} \r)^2  }
\end{eqnarray}
Note that our choice of the Hamiltonian corresponds to choosing $\tau$ as the relevant time coordinate; in other words, $H=-\bm p \bm \cdot \bm \partial_{\tau}$. Since the metric is $\eta_{ab}$ all along the trajectory in FNC, this Hamiltonian correctly accounts for the additional ``gravity" field which would appear for particles located away from the trajectory, since the zero of gravity potential is (by construction) at the trajectory. 
%To illustrate the point, compare this with the case of a static box in Schwarzschild, with its center at some $r=r_0$. From some simple manipulations, it is easy to see that the Hamiltonian as constructed above becomes, in the non-relativistic limit, $H-mc^2=p^2/2m + (GM/r_0^2) m y$, which is the correct expression when the gravity potential at $r=r_0$ is chosen to be zero. 
This definition is also equivalent to the so called ``energy at infinity" defined in spacetimes with a timelike Killing vector; in this sense, the role of $\bm \ub=\bm \partial_{\tau}$ above is the same as that of the Killing vector when available. These conceptual points are also significant for what we mean by temperature $T=\beta^{-1}$. What $\beta^{-1}$ represents is the temperature of a reservoir (say, a heating element) at the center of the box, which is our reference point. This $\beta$ is therefore constant, unlike the Tolman temperature which is \textit{defined} with respect to local energy of the particles, and hence must be multiplied by a red-shift factor. The above discussion is important to correctly describe a thermal equilibrium state as a Gibbs state in a curved spacetime. 
%Another way to emphasize the physical aspect of this discussion is that, since heat has weight, a quantity of heat flowing from one point to another point at a different gravity potential will be changed in content by its gravitational potential energy, and hence the correct temperature assigned at a higher potential is lower than the temperature at a lower potential by the Tolman redshift factor. 

%Finally, it is worth mentioning that a rigorous analysis of Dirac equation in FNC, done long back by Parker \cite{parker-fnc}, yields, in the non-relativistic limit, the same Hamiltonian as what we shall obtain below from our definition.

%It might appear that since the canonical partition function depends only the combination $\beta E$, choosing the local temperature and local energy would give the same result. This is incorrect, and the subtlety lies in calculating the energy spectrum using a quantum mechanical wave equation. This spectrum depends on the choice of energy at the classical level, and hence the final results for the partition function would be very different if one were to choose local energy from the outset. The correct time coordinate in our case is provided by $\bm \ub$, since the boundary conditions on quantum mechanical wavefunction is provided by the walls of the box, which are constructed with respect to $\bm \ub$. %Had one chosen $\bm \ub = (-g_{00})^{-1/2} \bm \ub$ corresponding to local energy, there would be 

In the non-relativistic limit, the Hamiltonian reduces to
\begin{widetext}
\begin{eqnarray}
H - m c^2 = \frac{p^2}{2 m} \l( 1 + R_{{\hat 0} \mu {\hat 0} \nu} y^{\mu} y^{\nu} \r) + \frac{1}{2} m c^2 R_{{\hat 0} \mu {\hat 0} \nu} y^{\mu} y^{\nu} + \frac{2}{3} R_{{\hat 0} \nu \mu \rho} y^{\nu} y^\rho c p_{\mu} - \frac{1}{6} R_{\mu \rho \nu \sigma} y^\rho y^\sigma \frac{p_{\mu} p_{\nu}}{2 m}
\label{eq:hamiltonian-full}
\end{eqnarray}
\end{widetext}
Further, in the non-relativistic limit $p \ll mc$ and terms which go as $\mathcal{R} y^2 \times (p/mc)^2$ are second order of smallness, and therefore can be ignored. 
%Moreover, it follows from Einstein field equations that $\mathcal{R} \sim G \rho_m/c^2 $ where $\rho_m$ is mass density. 
Using all this, it is easy to see that, in the $c \rightarrow \infty$ limit, we are finally left with
\begin{eqnarray}
H - m c^2 = \frac{p^2}{2 m} + \frac{1}{2} m c^2 R_{{\hat 0} \mu {\hat 0} \nu} y^{\mu} y^{\nu} 
\end{eqnarray}
Incidentally, this expression matches with the non-relativistic limit of Dirac Hamiltonian obtained by Parker (see Eq.~(9.13) in \cite{parker-fnc}), which provides further separate support for the arguments leading to it. In what follows, we shall ignore the time dependence carried by curvature components. This is a reasonable assumption if time scale on which curvature changes, $\mathcal{R}/\mathcal{\dot R}$, is much larger than typical time scale associated with the gas; we expect this to be the case at high enough temperature. 

The energy eigenvalues can be easily found using first order perturbation theory, with the unperturbed eigenfunctions being the standard ones
$\overset{\circ}{\psi}_{\{n_i\}}(\mathrm{\bf y}) = \sqrt{8/V} \prod \limits_{i=1,3} 
\l\{ 
\sin{ \l[2 n_i \pi y_i/L_i \r] } ; \cos{ \l[(2 n_i-1) \pi y_i/L_i \r] }
\r\}
$.
%%\begin{widetext}
%%\begin{eqnarray}
%%\overset{\circ}{\psi}_{\{n_i\}}(\mathrm{\bf y}) = \sqrt{\frac{8}{V}} \Biggl\{  \;  \prod \limits_{i=1,3} \sin{ \l[ \frac{2 n_i \pi}{L_i} y_i \r] }  \;\;\; ; \;\;\;     \prod \limits_{i=1,3} \cos{ \l[ \frac{(2 n_i-1) \pi}{L_i} y_i \r] }  \;  \Biggl\}
%%\end{eqnarray}
%%\end{widetext}
%%corresponding to different parities. 
Here $V=L_x L_y L_z$ is the box volume and $n_i \in [1,\infty)$. We assume that the sides of the box are incommensurable, so that non-degenerate perturbation theory can be used. Then one obtains %\footnote{Note that, had one kept the third term in (\ref{eq:hamiltonian-full}) despite it being sub-dominant, one would have had to deal with ordering ambiguity due to momentum dependence. Interestingly, however, the contribution of this term to perturbed energy eigenvalues seem to vanish for all possible choices of factor ordering!}
\begin{eqnarray}
E_{ \{n_i\} } &=& \overset{\circ}{E}_{ \{n_i\} } + \frac{1}{24} m c^2 \sum \limits_{i=1,3} R_{{\hat 0} i {\hat 0} i} L_i^2 \l( 1 - \frac{6}{\pi^2 n_i^2} \r)
\nonumber \\ 
\overset{\circ}{E}_{ \{n_i\} } &=& \frac{\hbar^2 \pi^2}{2 m} \sum \limits_{i=1,3} \frac{n_i^2}{L_i^2}
\label{eq:pert-eev}
\end{eqnarray}
The partition function for $N$ particles, assuming Boltzmann statistics, is $Z=z^N/N!$ where $z$ is the one particle partition function. This can be calculated by approximating intermediate sums as integrals, and assuming $\lambda^3/V \ll 1$, where $\lambda = h/\sqrt{2 \pi m k T}$ is the thermal de Broglie wavelength. First, define the functions
\begin{eqnarray}
p(s) &=& \sum \limits_{n=1}^{\infty} \frac{ \exp{ \l[- (\pi/4) s^2 n^2 \r]} }{ n^2 }
\nn \\
q(s) &=& s p(s)
\end{eqnarray}
The behaviour of $q(s)$ can be deduced by properly approximating sums with integrals. However, as we shall see, the final results can be stated simply in terms of $q(s)$ and its derivatives at $s=0$, so all that is required is that $q(s)$ be analytic at $s=0$. Also, in what follows, terms of the form ``$R \lambda^3/L$" have been ignored, since they are small compared to the terms retained by factors of $\lambda/L$. The only sub-dominant term retained below is the one which is independent of box dimensions $L_i$ and mass $m$ of constituent particles, and hence expected to be of some fundamental significance -- indeed, it is this term which will turn out to have an interesting form. 

The canonical partition function turns out to be
{\small{
\begin{eqnarray}
\ln{\l(Z/Z_{\rm F}\r)} &=& \beta N m c^2 \Biggl\{ 
\l[ \frac{q''(0)}{8 \pi^2} \r]  R_{{\hat 0 \hat 0}} \lambda^2
\nn \\
&+& \sum \limits_{i=1,3} \Biggl( \l[ \frac{q'(0)}{4 \pi^2} \r] R_{{\hat 0} i {\hat 0} i} L_i \lambda - \frac{1}{24} R_{{\hat 0} i {\hat 0} i} L_i^2 \Biggl) 
\Biggl\}
\label{eq:finalZN}
\end{eqnarray}
}} 
where $\ln Z_{\mathrm{F}}=\ln (V^N \lambda^{-3N}/N!)$ is the flat space expression. To put the final results in a neat form, it is convenient to introduce the following

\vspace{0.2cm}

\textsc{Definitions}: %DEFINITIONS:
\begin{itemize}  \itemsep=0.051cm

%\vspace{0.2cm}

\item[] $\Lambda=\beta \hbar c$ which is a length scale independent of mass $m$ (unlike $\lambda$), and is therefore more fundamental

%\vspace{0.2cm}

\item[] $\mathcal{R}_i=R_{{\hat 0} i {\hat 0} i}$ and $\delta_i = L_i / \lambda \; (\gg 1)$

%\vspace{0.2cm}

\item[] $c_1 = -q''(0)/(2\pi)$ and $c_2 = q'(0)/(2\pi)$; numerical values of these are not relevant, but can be shown to be $c_1=1/2$, $c_2=\pi/12$

%\vspace{0.2cm}

\item[] $U_{\mathrm{corr}} = U-U_{\mathrm{F}}$ and $S_{\mathrm{corr}} = S-S_{\mathrm{F}} $, where $U_{\mathrm{F}}=3N/2\beta$ and 
$S_{\mathrm{F}} = 3N/2 + N \ln \l(e V/N \lambda^3\r)$ are standard flat space expressions

\vspace{0.1cm}

\end{itemize}
It is now straightforward to use standard definitions $U=-\partial_{\beta} \ln Z$ and $S=\ln Z + \beta U$ to evaluate $U_{\mathrm{corr}}$, $S_{\mathrm{corr}}$ and heat capacity at constant volume, $C_V=-\beta^2 \partial_{\beta} U = 3N/2 + C_{V \mathrm{corr} }$. We obtain
%\begin{widetext}
%\begin{eqnarray}
%\frac{1}{\l( 2 \pi N \r)} \;\; \beta U_{\mathrm{corr}} &=& c_1  R_{{\hat 0 \hat 0}} \Lambda^2 
%- c_2 \Lambda^2 \sum \limits_{i=1,3} \mathcal{R}_i \delta_i \l(  {36 c_2} - \delta_i \r) /24 c_2
%\nonumber \\
%\nonumber \\
%\frac{1}{\l( 2\pi N \r)} \;\; 2 S_{\mathrm{corr}} &=& c_1 R_{{\hat 0 \hat 0}} \Lambda^2  - c_2 \Lambda^2 \sum \limits_{i=1,3}  \mathcal{R}_i \delta_i
%\nonumber \\
%\nonumber \\
%\frac{C_V}{N} - \frac{3}{2} &=& c_1 R_{{\hat 0 \hat 0}} \Lambda^2  - \frac{3}{4}  c_2 \Lambda^2 \sum \limits_{i=1,3}
% \mathcal{R}_i \delta_i
%\end{eqnarray} 
%\end{widetext}
\begin{widetext}
\begin{eqnarray}
{2 S_{\mathrm{corr}}}/{N} &=& + c_1 R_{{\hat 0 \hat 0}} \Lambda^2  \; -\; c_2 \Lambda^2 \sum \limits_{i=1,3}  \mathcal{R}_i \delta_i \;+\; O(\delta_i^{-1})
\nonumber \\
\nonumber \\
{\beta U_{\mathrm{corr}}}/{N} &=& + c_1  R_{{\hat 0 \hat 0}} \Lambda^2 
%\; -\;  \l({1}/{24}\r) c_2 \Lambda^2 \sum \limits_{i=1,3} \mathcal{R}_i \delta_i \l(  {36} - 2 \pi c_2^{-1} \delta_i \r)
\; -\;  \l({3}/{2}\r)  c_2 \Lambda^2 \sum \limits_{i=1,3} \mathcal{R}_i \delta_i 
%\;+\; \l(\pi/12\r) \Lambda^2 \sum \limits_{i=1,3} \mathcal{R}_i \delta_i^2
\;+\; \beta \l(1/24\r) m c^2 \sum \limits_{i=1,3} \mathcal{R}_i L_i^2 \;+\; O(\delta_i^{-1})
\nonumber \\
\nonumber \\
{C_{V \mathrm{corr} }}/{N} &=& - c_1 R_{{\hat 0 \hat 0}} \Lambda^2  \; +\;  \l({3}/{4}\r)  c_2 \Lambda^2 \sum \limits_{i=1,3}
 \mathcal{R}_i \delta_i \;+\; O(\delta_i^{-1})
 \label{eq:final-expr}
\end{eqnarray} 
\end{widetext}
%which are correct to $O(\delta_i^{-1})$. 
Note that the third term on RHS of the second expression above, for mean energy per particle $U_{\mathrm{corr}}/{N}$, is simply due to the constant term in perturbed energy eigenvalues (see Eq.~(\ref{eq:pert-eev})). 

These are the final expressions we wanted to derive, with several features worth highlighting:
 
\begin{enumerate} \itemsep=0.5cm
%\vspace{0.5cm}
%\item if all sides of the box are of same length, $L_i = L \; \forall i$, then the results depend only on $\sum \limits_{i=1,3} R_{{\hat 0} i {\hat 0} i} = R_{{\hat 0 \hat 0}}=R_{ab} \ub^a \ub^b$. This, however, is not strictly allowed since then one must use degenerate perturbation theory.

\item The $L_i$ and $m$ independent part of the corrections, which we denote by $\Xi$, depend only on $\Xi = R_{{\hat 0} {\hat 0}} \Lambda^2$ with $R_{{\hat 0} {\hat 0}} = R_{ab} \; \ub^a \; \ub^b$.

\item Further, $S_{\Xi}$ and $U_{\Xi}$ satisfy
 \begin{eqnarray}
S_{\Xi} = \frac{1}{2} \beta U_{\Xi}
\end{eqnarray}
where $S_{\Xi}=\l( Nc_1/2 \r) \Xi$ and so on. This is Euler relation of homogeneity two, and is well known from black hole thermodynamics; in particular, black hole horizons have temperature, entropy and energy which also satisfy $S=(1/2) \beta U$, with $U$ being the Komar energy.
\footnote{This relation is quite general, and holds for charged, rotating black holes in higher dimensions as well \cite{bibhas-smarr}. Moreover, it also extends beyond Einstein gravity, to static horizons in Lanczos-Lovelock models \cite{paddy-equip}.
}

In fact, such a relation between $S$ and $U$ acquires importance in the {\it emergent gravity} viewpoint, since it can be interpreted as {\it equipartition law} for energy of microscopic degrees of freedom associated with spacetime horizons \cite{paddy-equip}. It is therefore quite intriguing that the $\Xi$ correction terms, which appear due to quantum mechanics and are independent of system details, have features in common with thermal features of spacetime horizons.

The relevance of such Euler relation and area scaling of entropy for self-gravitating systems has also been emphasized in \cite{oppenheim}. 

%\item It is easy to show that $q''(0)=-\pi<0$; hence, if matter satisfies the Strong Energy Condition, then $R_{ab}\ub^a \ub^b \geq 0$, the first term on RHS represents the \textit{minimum} correction to mean energy $U$ and entropy $S$, and it is independent of box dimensions and mass of particles. Therefore, this term is the minimal correction to entropy and energy independent of any details of the box or its contents, but depending only on temperature and background curvature.

\item The $\Xi$ contribution to specific heat is {\it negative} if the strong-energy condition ($R_{{\hat 0} {\hat 0}} \geq 0$) holds.
\end{enumerate}
Incidentally, note that it is possible to incorporate finite size effects in the calculation rather simply by approximating $\sum \limits_{n=1}^{\infty} \exp{(-\alpha n^2)} \sim (  \sqrt{{\pi}/{\alpha}} - 1 )/2 $ instead of just $\sqrt{{\pi}/{\alpha}}/2 $ as is usually done; if one does this, terms involving surface area $A=2(L_x L_y + L_y L_z + L_z L_x)$ of the box gets added to various quantities such as $S/N$, $U/N$ etc. in Eqs.~(\ref{eq:final-expr}). These additional terms are of the form $c_0 A \lambda/V = 2 c_0 \sum_i \delta^{-1}_i$ where $c_0$ is a number less than unity (see \cite{molina}). Therefore, to compare these surface corrections with the curvature corrections, one would have to keep the (lower order) $\delta^{-1}_i$ curvature terms in Eqs.~(\ref{eq:final-expr}), and then the comparison of curvature and surface corrections would be determined by relative magnitudes of $c_0$ and $\mathcal R_i \Lambda^2$ (both of which are less than unity).

Finally, let us point out an important fact concerning the flat spacetime limit of the result. The curvature terms which appear in FNC refer to the background spacetime, say $\mathcal{S}$, which must be evaluated on the trajectory $\bm \ub$ which is a geodesic of $\mathcal{S}$. However, since our system has finite energy, for consistency we must consider a geodesic in the spacetime comprising of background {\it and} the box contents, $\mathcal{S}+\mathcal{B}$, where $\mathcal{B}$ is the perturbation to $\mathcal{S}$ caused by the box contents. Therefore, even when $\mathcal{S}$ is flat, there will always be some contribution from curvature produced by energy (rest + thermal) in the box.

%%%%%%%%%%%%%%%%%%%%%%%%%%%%%%%%%%%%%%%%%%%%%%%%%%%%%%%%%%%%%%%%%%%%%%%%%%%%%%%%%%%%%%%%%%%%%
\section{Discussion and Implications}%\vspace{0.2cm}
%\noindent \textbf{\textit{Discussion and Implications}}: 
It is well known that thermal behaviour of systems with long-range interactions exhibit several peculiar features, and their statistical analysis requires considerable amount of care \cite{lynden-bell-paddy}. Some of these features, such as negative specific heat and unequal ``local" temperature in thermal equilibrium, became more widely known since the discovery of thermodynamic aspects of black holes. However, these same aspects, which have their origin in quantum field theory in non-trivial coordinate systems, also indicate that one must take a closer look at behaviour of {\it standard} thermal systems in an external gravitational field and in presence of spacetime horizons. For example, temperature and entropy associated with local acceleration horizons are observer dependent, and one must therefore try to find a natural way to incorporate these inevitable features in conventional thermodynamics. These and related issues form the main motivation for the analysis in this note. 

Let us first summarize the main steps of the calculation. We defined the non-relativistic Hamiltonian in FNC, and then found its energy eigenvalues by using first order perturbation theory. It must be emphasized that the role of the box in this calculation is rather subtle, and it does {\it not} lead just to a pure finite size effect; indeed, the $\Xi$ term is independent of $L_i$ and $m$. We must also emphasize that the interesting features we have highlighted are shown only by the Ricci corrections, and not by the full expressions. The connection with thermodynamic features of black holes would require us to probe deeper into the physical significance of pure Ricci corrections
\footnote{Note that Ricci tensor does play the key role in defining, for example, entropy in Einstein theory. Specifically, the Noether current of diffeomorphism invariance which gives the entropy is proportional to Ricci in Einstein theory.}
. In this context, it is also worth mentioning that the time scale $t_o=\Lambda/c=\hbar/kT$ appearing in the Ricci term has recently been discussed by Haggard and Rovelli \cite{haggard-rovelli} as the average time a system in thermal equilibrium at temperature $T$ takes to move from a state to the next distinguishable state, thereby making this time step universal and independent of any details of the system other than its temperature.

One context in which the result might be relevant is the so called generalized second law (GSL), which was formulated by Bekenstein \cite{bekenstein-entropy} based on the fact that horizons have entropy (proportional to area). GSL states that the total entropy of matter and horizon (when present), never decreases. {\it Classically}, the minimum increase in entropy of horizon, when a particle (or a system) falls across it, is actually zero -- it acquires a non-zero value only when quantum effects are taken into account, using which examples can be given (as Bekenstein did) which conform with GSL. In our case also, quantum mechanics yields an additional term proportional to $\Xi=R_{{\hat 0}{\hat 0}} \Lambda^2$ (besides other terms depending on details of the system) in the expression for entropy $S$ of the system. However, note that the curvature correction in Eq.~(\ref{eq:pert-eev}) does {\it not} have $\hbar$ in it -- the only non-trivial aspect at this level is the dependence of energy eigenvalues on $1/n_i^2$. Nevertheless, quantum nature of the result is evident in the $\Xi$ corrections which are $O(\hbar^2)$ since $\Lambda=O(\hbar)$. It would be interesting to investigate how (if at all) does this affect the analyses related to GSL.

It is, of course, important to generalize the result presented here to different systems, and also to the relativistic case. However, the fact that the relevant term depends only on the length scale $\Lambda=\beta \hbar c$ suggests that the result will survive in the relativistic limit (since it does not depend on mass). \footnote{In fact, a close look at the results in \cite{parker-fnc} indicates that relativistic corrections produce {\it qualitative} changes only in $l \neq 0$ energy levels of Hydrogen atom, which strengthens this belief further.} A rigorous analysis of other systems (such as harmonic oscillator) confined in a box is complicated by the fact that even unperturbed energy eigenvalues and eigenfunctions are not known in analytic form. One hopes to arrive at a better understanding of deeper physical implications of the result provided such technical issues could be overcome. Incidentally, note that the time scale $\Delta t = \Lambda/c=\hbar/kT$ is the time uncertainty associated with thermal fluctuations of energy $kT$. There is a hint here of interplay between thermal and quantum fluctuations, but again, more work is needed to understand the underlying physics better.

%Finally, we end with a speculation related to black hole entropy. If we \textit{assume} that the result, $S_{\Xi}/N=(1/2) c_1 R_{{\widehat 0}{\widehat 0}} \Lambda^2$, is generic to \textit{all} systems, and apply it to black holes (i.e., use Hawking temperature for $\beta$), we will obtain $S_{\Xi}=(2 \pi N R_{{\widehat 0}{\widehat 0}}) \pi r_H^2$ where $r_H$ is horizon radius. We see that the proportionality constant between entropy and area depends on $N$ and $R_{{\widehat 0}{\widehat 0}}$, the physical interpretation of neither of which is clear for black holes.
\vspace{0.2cm}
I thank T Padmanabhan for discussion. Major part of this research was funded by {National Science \& Engineering Research Council} (NSERC) of Canada, and Atlantic Association for Research in the Mathematical Sciences (AARMS).

\appendix

\vspace{0.2cm}
\section{An outline of steps leading to Eq.~(\ref{eq:finalZN})}

We here briefly outline the steps needed to arrive at Eq.~(\ref{eq:finalZN}). Let $E=E_0+E_1$ with
\begin{eqnarray}
E_0 &=& \frac{\hbar^2 \pi^2}{2 m} \sum \limits_{i=1}^{3} \frac{n_i^2}{L_i^2}
\;\; ; \;\;\;\;\;\;
E_1 = \sum \limits_{i=1}^{3} A_i \l( 1 - \frac{6}{\pi^2 n_i^2} \r) 
\nn
\end{eqnarray}
where $A_i=(1/24) mc^2 R_{0i0i} L_i^2$.
The one-particle partition function is then
\begin{eqnarray}
z &=& \sum \limits_{\{n_i\}} \exp{ \l[- \beta \l( E_0 + E_1 \r) \r] }
\nn \\
&\approx & \sum \limits_{\{n_i\}} \exp{\l[ - \beta E_0 \r]} \; - \beta \sum \limits_{\{n_i\}} E_1 \exp{ \l[- \beta E_0 \r]} 
%+ O(R_{abcd}^2, \nabla R_{abcd})
\label{eq:fullz}
\end{eqnarray}
to $O(R_{abcd}^2, \nabla R_{abcd})$. The first term on RHS is standard textbook expression, and 
%can be evaluated using
%\begin{eqnarray}
%z_0 \equiv \sum \limits_{n=1}^{\infty} \exp{ - \l( \frac{\beta \hbar^2 \pi^2}{2 m} \frac{n^2}{L_i^2} \r)} \approx \frac{L_i}{\lambda} = \delta_i \;\; (\mathrm{say})
%\end{eqnarray}
%in the approximation $\delta_i \gg 1$. This gives the standard result $z_0 = {V}/{\lambda^3}$. The second term on RHS of Eq.~(\ref{eq:fullz}) can be expanded out as
evaluates to ${V}/{\lambda^3} \equiv z_0$. The second term on RHS of Eq.~(\ref{eq:fullz}) can be expanded out as
%\begin{widetext}
\begin{eqnarray}
\sum \limits_{\{n_i\}} E_1 \exp{ \l[- \beta E_0 \r]} &=& z_0 \sum \limits_{i=1}^{3} \Biggl[ A_i \; - \frac{6}{\pi^2} \frac{A_i p_i}{\delta_i} \Biggl]
\end{eqnarray}
%\end{widetext}
where $p_i \equiv p(s)|_{s=\lambda/L_i}$ and $p_i/\delta_i=p_i\lambda/L_i=q_i \equiv q(s)|_{s=\lambda/L_i}$. Substituting in Eq.~(\ref{eq:fullz}), and defining $\mathcal R_i = R_{0i0i}$ for shorthand, we get
\begin{eqnarray}
z &=& z_0 \Biggl[ 1 - \frac{\beta mc^2 }{24} \sum \limits_{i=1}^{3} \mathcal R_i L_i^2 + \frac{\beta mc^2}{4 \pi^2} \sum \limits_{i=1}^{3} \mathcal R_i L_i^2 q_i \Biggl]
\nn
\end{eqnarray}
The final expression for the full partition function can now be obtained easily by expanding $q_i$'s for small $\lambda/L_i$ in Taylor series ($q(0)=0$ is easily verified): $q_i = q'(0) \lambda/L_i + (q''(0)/2)(\lambda/L_i)^2 + O(\lambda^3/L_i^3)$. The coefficients $q''(0)$ and $q'(0)$, which are eventually responsible for the terms independent and linear in $L_i$ respectively in Eq.~(\ref{eq:finalZN}), can be easily evaluated using: $p(0)=\pi^2/6$, $p'(0)=-(\pi/2)[sG(s)]_{s=0}=-\pi/2$ where $G(s)$ is the Gauss sum $G(s) = \sum_{n=1}^{\infty} \exp{\l[- (\pi/4) s^2 n^2 \r]}$ with well known small $s$ expansion, $G(s) \approx s^{-1} - 1/2 + O(s^2)$. Similarly one can find leading term in $p''(s)$ as well. Putting everything together, we eventually recover $q'(0)=\pi^2/6$ and $q''(0)=-\pi$, which give the values of $c_1, c_2$ quoted in the text. It is easy to see that our assumptions of the gas being non-relativistic, and $\delta_i \gg 1 ~ \forall i$, imply that the range of temperature over which the entire calculation is valid is: ${\hbar^2}/{m L^2} \ll kT \ll mc^2$ (where $L={\rm min}\{L_i\}$ and factors of $2 \pi$ etc have been ignored). The ratio of upper to lower limits here is: $(L/\lambda_c)^2$, where $\lambda_c=\hbar/mc$ is the Compton wavelength. For ordinary systems, this provides a considerable range over which $kT$ can vary while satisfying all the approximations made in the paper.

\vspace{0.2cm}
\section{Average Pressure and Equation of State}

An important observation can also be made regarding average pressure of the gas. Pressure on the $i-$th face can be defined as $\beta P_i= ({ L_i}/{V}) {\partial \ln Z}/{\partial L_i}$ since ${V}/{L_i}=A_{\perp i}$ is the transverse area of the face. If one defines the correction to average pressure by $P=(1/3) \sum_i P_i = P_{\mathrm{corr}} + P_{\mathrm{F}}$, where $\beta P_{\mathrm{F}}=N/V$ is the flat space expression, then one can easily show that
\begin{eqnarray}
\beta P_{\mathrm{corr}} = \frac{4}{3} \frac{1}{V} \l( S_{\mathrm{corr}} - \frac{1}{2} \beta U_{\mathrm{corr}}  \r)
\end{eqnarray}
As mentioned above, $ S_{\Xi} = \frac{1}{2} \beta U_{\Xi}$. Hence, what the above expression says is that, the average pressure is contributed by those parts of $ S_{\mathrm{corr}}$ and $U_{\mathrm{corr}} $ which \textit{do not} satisfy $S=(1/2) \beta U$.
%\footnote{
%In flat spacetime
%$\beta P_{\mathrm{F}} V = \frac{1}{\ln{\l[e V/(N \lambda^3)\r] }} \l( S_{\mathrm{F}} - \beta U_{\mathrm{F}} \r)$ where the prefactor on the RHS is related to chemical potential of the ideal gas.
%} 
Using the above result, one can write down the correction to equation of state of the gas:
\begin{eqnarray}
\frac{\beta P V}{N} = 1 - \l(\frac{\Lambda}{6}\r)^2 \sum \limits_{i=1,3} \mathcal{R}_i \delta_i \l( 12 c_2 + 2 \pi \delta_i \r)
\end{eqnarray}
%Note that the final pressure {\it does} depend on geometric features of the box. In fact, it might be possible to interpret the second term in the round brackets on RHS as additional pressure due to tidal forces acting on the box. Quantum mechanics adds to this the $c_2$ piece (note that $\Lambda^2 \delta_i \sim O(\hbar^1)$ while $\Lambda^2 \delta_i^2 \sim O(\hbar^0)$). 

%\vspace{0.2cm}
%\section{A note on approximations}
%
%The results in this paper have been derived assuming: (i) the gas is non-relativistic, i.e., $kT \ll mc^2$, and (ii) $\delta_i \gg 1 ~ \forall i$. The latter implies $k T \gg \hbar^2 / m L^2$ (where $L$ is the length of the smallest side of the box, and factors of $2 \pi$ etc have been ignored). Combined with (i), this gives the range of temperature over which the entire calculation is valid:
%\begin{eqnarray}
%\frac{\hbar^2}{m L^2} \ll kT \ll mc^2
%\end{eqnarray}
%The ratio of the upper to lower limit here is: $mc^2/(\hbar^2 / m L^2) = (L/\lambda_c)^2$, where $\lambda_c=\hbar/mc$ is the Compton wavelength. For realistic systems, this provides a considerable range over which $kT$ can vary while satisfying all the approximations made in the paper.

%%%%%%%%%%%%%%%%%%%%%%%%%%%%%%%%%%%%%%%%%%%%%%%%%%%%%%%%%%%%%%%%%%%%%%%%%%%%%%%%%%%%%%%%%%

%%%%%%%%%%%%%%%%%%%%%%%%%%%%%%%%%%%%%%%%%%%%%%%%%%%%%%%%%%%%%%%%%%%%%%%%%%%%%%%%%%%%%%%%%%

%%%%%%%%%%%%%%%%%%%%%%%%%%%%%%%%%%%%%%%%%%%%%%%%%%%%%%%%%%%%%%%%%%%%%%%%%%%%%%%%%%%%%%%%%%

\end{document}